# 1. BIG-BANG NUCLEOSYNTHESIS

Revised October 2013 by B.D. Fields, (Univ. of Illinois) P. Molaro (Trieste Observatory) and S. Sarkar (Univ. of Oxford & Niels Bohr Institute, Copenhagen).

Big-Bang nucleosynthesis (BBN) offers the deepest reliable probe of the early Universe, being based on well-understood Standard Model physics [1]. Predictions of the abundances of the light elements, D, $^3$He, $^4$He, and $^7$Li, synthesized at the end of the 'first three minutes', are in good overall agreement with the primordial abundances inferred from observational data, thus validating the standard hot Big-Bang cosmology (see [2–4] for reviews). This is particularly impressive given that these abundances span nine orders of magnitude – from $^4$He/H $\sim 0.08$ down to $^7$Li/H $\sim 10^{-10}$ (ratios by number). Thus BBN provides powerful constraints on possible deviations from the standard cosmology, and on new physics beyond the Standard Model [5–8].

## 1.1. Theory

The synthesis of the light elements is sensitive to physical conditions in the early radiation-dominated era at a temperature $T \sim 1$ MeV, corresponding to an age $t \sim 1$ s. At higher temperatures, weak interactions were in thermal equilibrium, thus fixing the ratio of the neutron and proton number densities to be $n/p = e^{-Q/T}$, where $Q = 1.293$ MeV is the neutron-proton mass difference. As the temperature dropped, the neutron-proton inter-conversion rate per nucleon, $\Gamma_{n\leftrightarrow p} \sim G_F^2 T^5$, fell faster than the Hubble expansion rate, $H \sim \sqrt{g_* G_N}\, T^2$, where $g_*$ counts the number of relativistic particle species determining the energy density in radiation (see 'Big Bang Cosmology' review). This resulted in departure from chemical equilibrium ('freeze-out') at $T_{\rm fr} \sim (g_* G_N/G_F^4)^{1/6} \simeq 1$ MeV. The neutron fraction at this time, $n/p = e^{-Q/T_{\rm fr}} \simeq 1/6$, is thus sensitive to every known physical interaction, since $Q$ is determined by both strong and electromagnetic interactions while $T_{\rm fr}$ depends on the weak as well as gravitational interactions. Moreover, the sensitivity to the Hubble expansion rate affords a probe of, *e.g.,* the number of relativistic neutrino species [9]. After freeze-out, the neutrons were free to $\beta$-decay, so the neutron fraction dropped to $n/p \simeq 1/7$ by the time nuclear reactions began. A simplified analytic model of freeze-out yields the $n/p$ ratio to an accuracy of $\sim 1\%$ [10,11].

The rates of these reactions depend on the density of baryons (strictly speaking, nucleons), which is usually expressed normalized to the relic blackbody photon density as $\eta \equiv n_{\rm b}/n_\gamma$. As we shall see, all the light-element abundances can be explained with $\eta_{10} \equiv \eta \times 10^{10}$ in the range 5.7–6.7 (95% CL). With $n_\gamma$ fixed by the present CMB temperature 2.7255 K (see 'Cosmic Microwave Background' review), this can be stated as the allowed range for the baryon mass density today, $\rho_{\rm b} = (3.9–4.6) \times 10^{-31}\ {\rm g\,cm}^{-3}$, or as the baryonic fraction of the critical density, $\Omega_{\rm b} = \rho_{\rm b}/\rho_{\rm crit} \simeq \eta_{10} h^{-2}/274 = (0.021–0.025) h^{-2}$, where $h \equiv H_0/100\ {\rm km\,s}^{-1}\,{\rm Mpc}^{-1}$ is the present Hubble parameter (see Cosmological Parameters review).

The nucleosynthesis chain begins with the formation of deuterium in the process $p(n, \gamma){\rm D}$. However, photo-dissociation by the high number density of photons delays production of deuterium (and other complex nuclei) until well after $T$ drops below the binding energy of deuterium, $\Delta_{\mathrm{D}} = 2.23$ MeV. The quantity $\eta^{-1}\mathrm{e}^{-\Delta_{\mathrm{D}}/T}$, *i.e.*, the number of photons per baryon above the deuterium photo-dissociation threshold, falls below unity at $T \simeq 0.1$ MeV; nuclei can then begin to form without being immediately photo-dissociated again. Only 2-body reactions, such as $\mathrm{D}(p,\gamma)^3\mathrm{He}$, $^3\mathrm{He}(\mathrm{D},p)^4\mathrm{He}$, are important because the density by this time has become rather low – comparable to that of air!

Nearly all neutrons end up bound in the most stable light element $^4\mathrm{He}$. Heavier nuclei do not form in any significant quantity both because of the absence of stable nuclei with mass number 5 or 8 (which impedes nucleosynthesis via $n^4\mathrm{He}$, $p^4\mathrm{He}$ or $^4\mathrm{He}^4\mathrm{He}$ reactions), and the large Coulomb barriers for reactions such as $^3\mathrm{He}(^4\mathrm{He},\gamma)^7\mathrm{Li}$ and $^3\mathrm{He}(^4\mathrm{He},\gamma)^7\mathrm{Be}$. Hence the primordial mass fraction of $^4\mathrm{He}$, $Y_{\mathrm{p}} \equiv \rho(^4\mathrm{He})/\rho_{\mathrm{b}}$, can be estimated by the simple counting argument

$$Y_{\mathrm{p}} = \frac{2\,(n/p)}{1 + n/p} \simeq 0.25 \; . \tag{1.1}$$

There is little sensitivity here to the actual nuclear reaction rates, which are, however, important in determining the other 'left-over' abundances: D and $^3\mathrm{He}$ at the level of a few times $10^{-5}$ by number relative to H, and $^7\mathrm{Li}/\mathrm{H}$ at the level of about $10^{-10}$ (when $\eta_{10}$ is in the range 1–10). These values can be understood in terms of approximate analytic arguments [11,12]. The experimental parameter most important in determining $Y_{\mathrm{p}}$ is the neutron lifetime, $\tau_n$, which normalizes (the inverse of) $\Gamma_{n\leftrightarrow p}$. Its value has recently been significantly revised downwards to $\tau_n = 880.0 \pm 0.9$ s (see $N$ Baryons Listing).

The elemental abundances shown in Fig. 1.1 as a function of $\eta_{10}$ were calculated [13] using an updated version [14] of the Wagoner code [1]; other versions [15–17] are publicly available. The $^4\mathrm{He}$ curve includes small corrections due to radiative processes at zero and finite temperatures [18], non-equilibrium neutrino heating during $e^\pm$ annihilation [19], and finite nucleon mass effects [20]; the range reflects primarily the $2\sigma$ uncertainty in the neutron lifetime. The spread in the curves for D, $^3\mathrm{He}$, and $^7\mathrm{Li}$ corresponds to the $2\sigma$ uncertainties in nuclear cross sections, as estimated by Monte Carlo methods [14,21–23]. The input nuclear data have been carefully reassessed [13, 23–27], leading to improved precision in the abundance predictions. In particular, the uncertainty in $^7\mathrm{Li}/\mathrm{H}$ at interesting values of $\eta$ has been reduced recently by a factor $\sim 2$, a consequence of a similar reduction in the error budget [28] for the dominant mass-7 production channel $^3\mathrm{He}(^4\mathrm{He},\gamma)^7\mathrm{Be}$. Polynomial fits to the predicted abundances and the error correlation matrix have been given [22,29]. The boxes in Fig. 1.1 show the observationally inferred primordial abundances with their associated uncertainties, as discussed below.

The nuclear reaction cross sections important for BBN have all been measured at the relevant energies. We will see, however, that recently there have been substantial advances in the precision of light element observations (e.g., D/H ) and in cosmological parameters (e.g., from *Planck*). This motivates corresponding improvement in BBN precision and thus in the key reaction cross sections. For example, it has been suggested [30] that $d(p,\gamma)^3\mathrm{He}$ measurements may suffer from systematic errors and be inferior to *ab initio* theory; if so, this could alter D/H abundances at a level that is now significant.

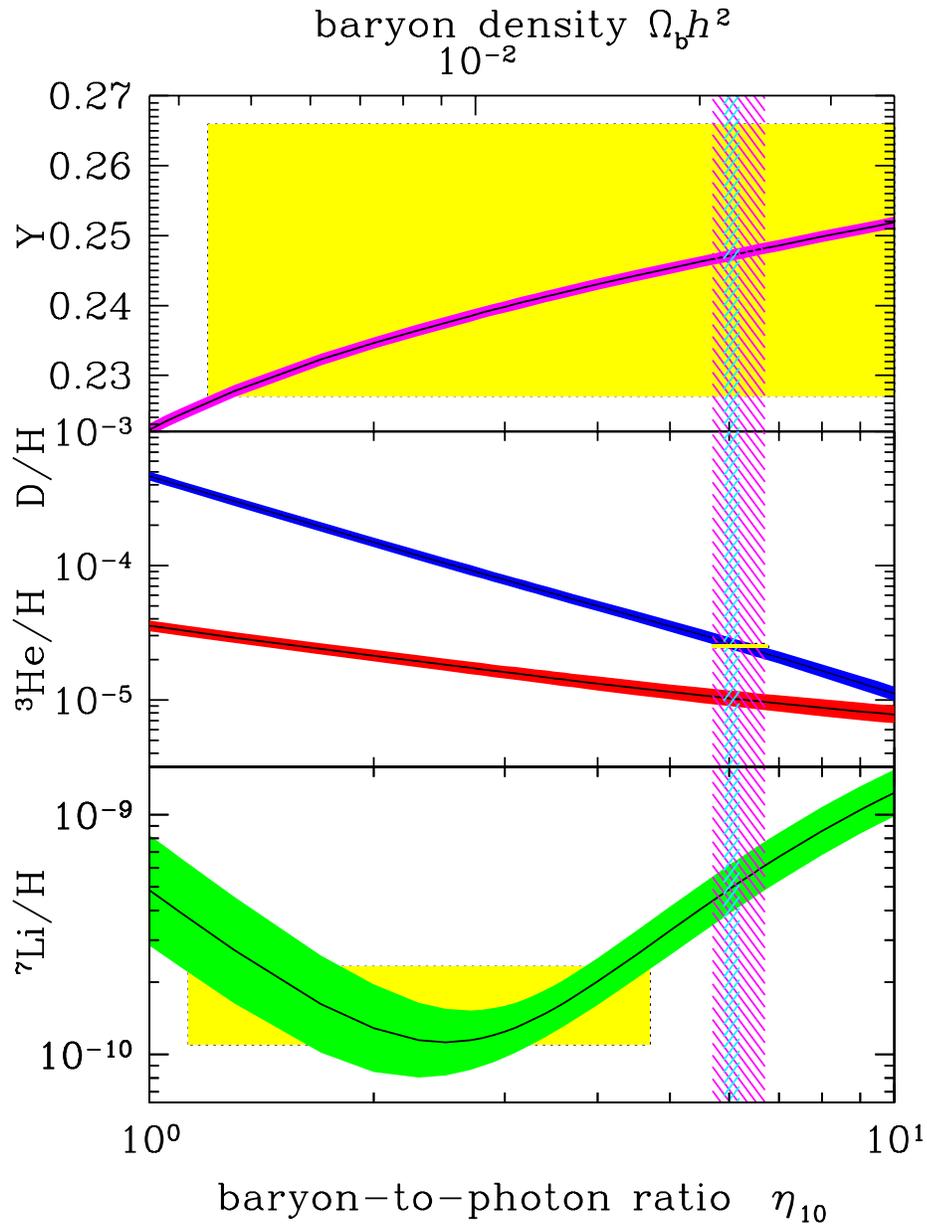

**Figure 1.1:** The abundances of $^4$He, D, $^3$He, and $^7$Li as predicted by the standard model of Big-Bang nucleosynthesis — the bands show the 95% CL range. Boxes indicate the observed light element abundances. The narrow vertical band indicates the CMB measure of the cosmic baryon density, while the wider band indicates the BBN concordance range (both at 95% CL).

## 1.2.  Light Element Abundances

BBN theory predicts the universal abundances of D, $^3$He, $^4$He, and $^7$Li, which are essentially fixed by $t \sim 180$ s. Abundances are, however, observed at much later epochs, after stellar nucleosynthesis has commenced. This produces heavy elements such as C, N, O, and Fe ('metals'), while the ejected remains of this stellar processing alters the light element abundances from their primordial values. Thus, one seeks astrophysical sites with low metal abundances, in order to measure light element abundances that are closer to primordial. For all of the light elements, systematic errors are the dominant limitation to the precision with which primordial abundances can be inferred.

BBN is the only significant source of deuterium, which is entirely destroyed when it is cycled into stars [31]. Thus, any detection provides a lower limit to primordial D/H, and an upper limit on $\eta_{10}$; for example, the local interstellar value of D/H $= (1.56 \pm 0.40) \times 10^{-5}$ [32] requires $\eta_{10} \leq 9$. The best proxy to the primordial value of D is its measure in distant and chemically unprocessed matter where stellar processing (astration) is minimal [31]. This has become possible with the advent of large telescopes, but after two decades of observational efforts we have only about a dozen determinations [33–41]. High-resolution spectra reveal the presence of D in high-redshift, low-metallicity quasar absorption systems via its isotope-shifted Lyman-$\alpha$ absorption features; these are, unfortunately, usually obscured by the Lyman-$\alpha$ forest. The available D measurements are performed in systems with metallicities from 0.1 to 0.001 Solar where no significant astration is expected [34]. In the best-measured systems, D/H shows no hint of correlation with metallicity, redshift or the hydrogen column density $N(\mathrm{H})$ ($= \int_{\mathrm{los}} n_{\mathrm{H}}\, ds$) integrated over the line-of-sight through the absorber. This is consistent with the measured D/H being representative of the primordial value.

The first measurements in 'damped' Lyman-$\alpha$ systems (DLAs: $N(\mathrm{H}) > 10^{20}$ cm$^{-2}$) [33,35] showed that D/H can be measured in this class of absorbers where the Lorentzian damping wings of Lyman-$\alpha$ and Lyman-$\beta$ (if relatively uncontaminated by Lyman-$\alpha$ clouds) provide a precise H column density. Subsequently DLA systems have been found that also show resolved higher members of the Lyman series. Systems with a particularly simple kinematic structure are desirable to avoid uncertainties with complex, only partially resolved components. Recently a DLA showing 13 resolved D I absorption lines has been analyzed together with 4 other suitable systems. This provides a strikingly improved precision over earlier work, with a weighted mean of $\log(\mathrm{D/H}) = -4.597 \pm 0.006$, corresponding to

$$\mathrm{D/H}|_{\mathrm{p}} = (2.53 \pm 0.04) \times 10^{-5}. \qquad (1.2)$$

D/H values in the Galaxy show an unexpected scatter of a factor of $\sim 2$ [42], with a bimodal distribution as well as an anti-correlation with metal abundances. This suggests that interstellar D not only suffers stellar astration but also partly resides in dust particles that evade gas-phase observations. This is supported by a measurement in the lower halo [43], which indicates that the Galactic D abundance has decreased by a factor of only $1.1 \pm 0.13$ since its formation. However in the DLA the dust content is apparently quite small; this is implied by the abundances of refractory elements such as Fe, Cr and Si, which are in nearly Solar proportions. Thus, the value derived in Eq. (1.2) appears safe against D depletion into dust grains.

The primordial $^4$He abundance is best determined through recombination emission lines of He and H in the most metal-poor extragalactic H II (ionized) regions, *viz.* blue compact galaxies. There is now a large body of data on $^4$He and CNO in these galaxies, with over 1000 such systems in the Sloan Digital Sky Survey alone [44,50]. These data confirm that the small stellar contribution to the helium abundance is positively correlated with metal production, so extrapolation to zero metallicity gives the primordial $^4$He abundance $Y_p$. However, H II regions are complex systems and several physical parameters enter in the He/H determination, notably the electron density and temperature, as well as reddening. Thus systematic effects dominate the uncertainties in the abundance determination [44,45]. In recent work that has accounted for the underlying $^4$He stellar absorption, and/or the newly derived values of the HeI-recombination and H-excitation-collisional coefficients, the $^4$He abundances have significantly increased. Some recent results are: $Y_p = 0.249 \pm 0.009$ [45]; $Y_p = 0.248 \pm 0.003$ [46]; $Y_p = 0.254 \pm 0.003$ or $0.252 \pm 0.001$ (depending on which set of He I emissivities are used) [47]; $Y_p = 0.2534 \pm 0.0083$ [48]; and $Y_p = 0.2465 \pm 0.0097$ [49]. In the first and last two determinations the quoted error is a conservative estimate of the systematic uncertainties. Ref. [50] used a subsample of 111 H II regions drawn from a sample of 1610 objects, as those providing $Y$ with an accuracy better than 3%; the linear regression in $Y - $O/H gives $Y_p = 0.254 \pm 0.0006$ (stat) $\pm 0.003$ (syst). Our recommended $^4$He abundance is

$$Y_p = 0.2465 \pm 0.0097, \qquad (1.3)$$

where we have adopted the result of the recent analysis with the most detailed error budget [49]. Note that it does not find a significant growth of $Y_p$ with O/H.

The CMB damping tail is sensitive to the primordial $^4$He abundance, and is independent from both BBN and local $^4$He measurements. [51]. Recent measurements yield $Y_p = 0.266 \pm 0.021$ (see Cosmic Microwave Background review) [52], *i.e.*, consistent with the H II region helium abundance determination.

As we will see in more detail below, the primordial abundance of $^7$Li now plays a central role in BBN, and possibly points to new physics. The systems best suited for Li observations are metal-poor (Pop II) stars in the spheroid of the Galaxy, which have metallicities going down to perhaps $10^{-5}$ of the solar value [53]. Observations have long shown [54–57] that Li does not vary significantly in Pop II stars with metallicities $\lesssim 1/30$ of Solar — the 'Spite plateau' [54]. However there are systematic uncertainties due to different techniques used to determine the physical parameters (*e.g.*, the temperature) of the stellar atmosphere in which the Li absorption line is formed. Different analyses and in some cases different stars and stellar systems (globular clusters), yield Li/H$|_p = (1.7 \pm 0.3) \times 10^{-10}$ [57], Li/H$|_p = (2.19 \pm 0.28) \times 10^{-10}$ [58], and Li/H$|_p = (1.86 \pm 0.23) \times 10^{-10}$ [59].

Recent observations find a puzzling drop in Li/H in metal-poor stars with [Fe/H] $\equiv \log_{10}[(\text{Fe/H})/(\text{Fe/H})_\odot] < -3.0$ [60,61,62]. In particular Li is not detected in the two metal poor dwarfs with metallicities of [Fe/H] $\lesssim -5$ (HE 1327−2326 and SDSS

J102915+172927 where Li/H $< 10^{-11}$) suggesting that something is depleting Li at very low metallicity [63]. Since these stars have small masses and are almost fully convective during pre-main sequence evolution, they could have burned some Li [64]. The same effect may produce the melting of the Li plateau at low metallicities [61,62] thus making quite uncertain any primordial Li value obtained by extrapolating to zero metallicity. Hence to estimate Li/H$|_\text{p}$ it is safer to consider stars with $-2.8 <$ [Fe/H] $< -1.5$ [62] which provides

$$\text{Li/H}|_\text{p} = (1.6 \pm 0.3) \times 10^{-10}. \tag{1.4}$$

Moreover, the Li in Pop II stars may have been partially destroyed due to mixing of the outer layers with the hotter interior [65]. Such processes can be constrained by observations of the fragile isotope $^6$Li [55], and by the absence of significant scatter in Li versus Fe [56]. Li depletion by a factor as large as $\sim 1.8$ has been suggested [66].

Stellar determination of Li abundances typically sum over both stable isotopes $^6$Li and $^7$Li. Recent high-precision measurements are sensitive to the tiny isotopic shift in Li absorption (which manifests itself in the shape of the blended, thermally broadened line) and indicate $^6$Li/$^7$Li $\leq 0.05$ [67,68], thus confirming that $^7$Li is dominant. A claim of a $^6$Li plateau (analogous to the $^7$Li plateau) has been made [67], suggesting a significant primordial $^6$Li abundance. This has, however, been challenged by new observations and analyses [69,70,68], which show that stellar convective motions can generate asymmetries in the line shape that mimic the presence of $^6$Li. Hence the deduced abundance ratio $^6$Li/$^7$Li $< 0.05$ in the best studied stars presently provides a robust upper limit on the $^6$Li abundance [68].

Turning to $^3$He, the only data available are from the Solar system and (high-metallicity) H II regions in our Galaxy [71]. This makes inferring the primordial abundance difficult, a problem compounded by the fact that stellar nucleosynthesis models for $^3$He are in conflict with observations [72]. Consequently, it is no longer appropriate to use $^3$He as a cosmological probe; instead, one might hope to turn the problem around and constrain stellar astrophysics using the predicted primordial $^3$He abundance [73].

## 1.3. Concordance, Dark Matter, and the CMB

We now use the observed light element abundances to test the theory. We first consider standard BBN, which is based on Standard Model physics alone, so $N_\nu = 3$ and the only free parameter is the baryon-to-photon ratio $\eta$. (The implications of BBN for physics beyond the Standard Model will be considered below, Section 1.5). Thus, any abundance measurement determines $\eta$, and additional measurements overconstrain the theory and thereby provide a consistency check.

While the $\eta$ ranges spanned by the boxes in Fig. 1.1 do not all overlap, they are all within a factor $\sim 2$ of each other. In particular, the lithium abundance corresponds to $\eta$ values that are inconsistent with that of the (now very precise) D/H abundance as well as the less-constraining $^4$He abundance. This discrepancy marks the "lithium problem". The problem could simply reflect difficulty in determining the primordial lithium abundance; or could hint at a more fundamental omission in the theory. The possibility that lithium reveals new physics is addressed in detail in the next section. If however we exclude the lithium constraint because its inferred abundance may suffer from systematic uncertainties, then D/H and $^4$He are in agreement. The concordant $\eta$ range is essentially that implied by D/H, namely

$$5.7 \leq \eta_{10} \leq 6.7 \ (95\% \ \text{CL}) \,. \qquad (1.5)$$

Despite the lithium problem, the overall concordance remains remarkable: using only well-established microphysics we can extrapolate back to $t \sim 1$ s to predict light element abundances spanning 9 orders of magnitude, in approximate agreement with observation. This is a major success for the standard cosmology, and inspires confidence in extrapolation back to still earlier times.

This concordance provides a measure of the baryon content:

$$0.021 \leq \Omega_{\text{b}} h^2 \leq 0.025 \ (95\% \ \text{CL}) \,, \qquad (1.6)$$

a result that plays a key role in our understanding of the matter budget of the Universe. First we note that $\Omega_{\text{b}} \ll 1$, *i.e.*, baryons cannot close the Universe [74]. Furthermore, the cosmic density of (optically) luminous matter is $\Omega_{\text{lum}} \simeq 0.0024 h^{-1}$ [75], so that $\Omega_{\text{b}} \gg \Omega_{\text{lum}}$: most baryons are optically dark, probably in the form of a diffuse intergalactic medium [76]. Finally, given that $\Omega_{\text{m}} \sim 0.3$ (see Dark Matter and Cosmological Parameters reviews), we infer that most matter in the Universe is not only dark, but also takes some non-baryonic (more precisely, non-nucleonic) form.

The BBN prediction for the cosmic baryon density can be tested through precision observations of CMB temperature fluctuations (see Cosmic Microwave Background review). One can determine $\eta$ from the amplitudes of the acoustic peaks in the CMB angular power spectrum [77], making it possible to compare two measures of $\eta$ using very different physics, at two widely separated epochs. In the standard cosmology, there is no change in $\eta$ between BBN and CMB decoupling, thus, a comparison of $\eta_{\text{BBN}}$ and $\eta_{\text{CMB}}$ is a key test. Agreement would endorse the standard picture, while disagreement could point to new physics during/between the BBN and CMB epochs.

The analysis described in the Cosmic Microwave Background review, based on recent *Planck* data, yields $\Omega_{\text{b}} h^2 = 0.02207 \pm 0.00027$ or $\eta_{10} = 6.047 \pm 0.074$ [52]. As shown in Fig. 1.1, this CMB estimate of the baryon density (narrow vertical band) is remarkably consistent with the BBN range quoted in Eq. (1.6) and thus in very good agreement with the value inferred from recent high-redshift D/H measurements [36] and $^4$He determinations; together these observations span diverse environments from redshifts $z = 1000$ to the present [78].

The precision determinations of the baryon density using the CMB motivates the use of this value as an input to BBN calculations. Within the context of the Standard Model, BBN then becomes a zero-parameter theory, and the light element abundances are completely determined to within the uncertainties in $\eta_{\text{CMB}}$ and the BBN theoretical errors. Comparison with the observed abundances then can be used to test the astrophysics of post-BBN light element evolution [79]. Alternatively, one can consider possible physics beyond the Standard Model (*e.g.*, which might change the expansion rate during BBN) and then use all of the abundances to test such models; this is discussed in Section 1.5.

## 1.4.   The Lithium Problem

As Fig. 1.1 shows, stellar Li/H measurements are inconsistent with the CMB (and D/H), given the error budgets we have quoted. Recent updates in nuclear cross sections and stellar abundance systematics *increase* the discrepancy to over $5\sigma$, depending on the stellar abundance analysis adopted [13].

The question then becomes pressing as to whether this mismatch comes from systematic errors in the observed abundances, and/or uncertainties in stellar astrophysics or nuclear inputs, or whether there might be new physics at work [8].   Nuclear inputs (cross sections) for BBN reactions are constrained by extensive laboratory measurements; to increase $^7$Be destruction requires enhancement of otherwise subdominant processes that can be attained by missed resonances in a few reactions such as $^7$Be$(d,p)2\alpha$ if the compound nuclear state properties are particularly favorable   [80].   However, experimental searches have now closed off the most promising of these cases [81],  making a "nuclear fix" increasingly unlikely.

Another conventional means to solve the lithium problem is by *in situ* destruction over the long lifetimes of the host halo stars. Stellar depletion mechanisms include diffusion, rotationally induced mixing, or pre-main-sequence depletion. These effects certainly occur, but to reduce lithium to the required levels generally requires some *ad hoc* mechanism and fine tuning of the initial stellar parameters [64,82]. A putative signature of diffusion has been reported for the globular clusters NGC 6397 and NGC 6752, where the 'turnoff' stars exhibit slightly lower (by $\sim 0.1$ dex) abundances of Fe II, Ti II, Sc II, Ca I and Mg I, than in more evolved stars [66,83]. General features of diffusive models are a dispersion in the Li abundances and a pronounced downturn in the Li abundances at the hot end of the Li plateau. Some extra turbulence needs to be invoked to limit diffusion in the hotter stars and to restore uniform Li abundance along the Spite plateau   [82].   In the framework of these models (and also assuming identical initial stellar rotation) depletion by at most a factor $\sim 1.8$ is conceivable [66,83].

As nuclear and astrophysical solutions to the lithium problem become increasingly constrained (even if difficult to rule out definitively), the possibility of new physics arises. Nucleosynthesis models in which the baryon-to-photon ratio is inhomogeneous can alter abundances for a given $\eta_{\mathrm{BBN}}$, but will overproduce $^7$Li [84].   Entropy generation by some non-standard process could have decreased $\eta$ between the BBN era and CMB decoupling, however the lack of spectral distortions in the CMB rules out any significant energy injection upto a redshift $z \sim 10^7$ [85].   The most intriguing resolution of the lithium problem thus involves new physics during BBN   [6–8].

We summarize the general features of such solutions here, and later consider examples in the context of specific particle physics models. Many proposed solutions introduce perturbations to light-element formation during BBN; while all element abundances may suffer perturbations, the interplay of $^7$Li and D is often the most important *i.e.* observations of D often provide the strongest constraints on the allowed perturbations to $^7$Li. In this connection it is important to note that the new, very precise determination of D/H [36] will significantly constrain the ability of such models to ameliorate or solve the lithium problem.

A well studied class of models invokes the injection of suprathermal hadronic or electromagnetic particles due to decays of dark matter particles. The effects are complex and depend on the nature of the decaying particles and their branchings and spectra. However, the models that most successfully solve the lithium problem generally feature non-thermal nucleons, which dissociate all light elements. Dissociation of even a small fraction of $^4$He introduces a large abundance of free neutrons, which quickly thermalize. The thermal neutrons drive the $^7$Be$(n, p)^7$Li conversion of $^7$Be. The resulting $^7$Li has a lower Coulomb barrier relative to $^7$Be and is readily destroyed via $^7$Li$(p, \alpha)^4$He [86,87]. But $^4$He dissociation also produces D directly and via nonthermal neutron $n(p, \gamma)d$ reactions; this introduces a tension between Li/H reduction and D/H enhancement.

Another important class of models retains the standard cosmic particle content, but changes their interactions via time variations in the fundamental constants [88]. Here too, the details are model-dependent, but scenarios that solve or alleviate the lithium problem often feature perturbations to the deuteron binding energy. A weaker D binding leads to the D bottleneck being overcome later, so that element formation commences at a lower temperature and lower density. This leads in turn to slower nuclear rates that freeze out earlier. The net result is a *higher* final D/H, due to less efficient processing into $^4$He, but also *lower* Li, due to suppressed production via $^3$He$(\alpha, \gamma)^7$Be.

The lithium problem remains an unresolved issue in BBN. Nevertheless, the remarkable concordance between the CMB and the D (as well as $^4$He) abundance, remains a non-trivial success, and provides constraints on the early Universe and particle physics.

## 1.5. Beyond the Standard Model

Given the simple physics underlying BBN, it is remarkable that it still provides the most effective test for the cosmological viability of ideas concerning physics beyond the Standard Model. Although baryogenesis and inflation must have occurred at higher temperatures in the early Universe, we do not as yet have 'standard models' for these, so BBN still marks the boundary between the established and the speculative in Big Bang cosmology. It might appear possible to push the boundary back to the quark-hadron transition at $T \sim \Lambda_{\rm QCD}$, or electroweak symmetry breaking at $T \sim 1/\sqrt{G_{\rm F}}$; however, so far no observable relics of these epochs have been identified, either theoretically or observationally. Thus, although the Standard Model provides a precise description of physics up to the Fermi scale, cosmology cannot be traced in detail before the BBN era.

Limits on new physics come mainly from the observational bounds on the $^4$He abundance. This is proportional to the $n/p$ ratio when the weak-interaction rate falls behind the Hubble expansion rate at $T_{\rm fr} \sim 1$ MeV. The presence of additional neutrino flavors (or of any other relativistic species) at this time increases $g_*$, hence the expansion rate, leading to a larger value of $T_{\rm fr}$, $n/p$, and therefore $Y_{\rm p}$ [9,89]. In the Standard Model at $T = 1$ MeV, $g_* = 5.5 + \frac{7}{4}N_\nu$, where $N_\nu$ is the *effective* number of (nearly) massless neutrino flavors (see Big Bang Cosmology review). The helium curves in Fig. 1.1 were computed taking $N_\nu = 3$; small corrections for non-equilibrium neutrino heating [19] are included in the thermal evolution and lead to an effective $N_\nu = 3.04$ compared to assuming instantaneous neutrino freezeout (see Big Bang Cosmology review). The computed $^4$He abundance scales as $\Delta Y_{\rm p} \simeq 0.013 \Delta N_\nu$ [10]. Clearly the central value for $N_\nu$ from BBN will depend on $\eta$, which is independently determined (with weaker sensitivity to $N_\nu$) by the adopted D or $^7$Li abundance. For example, if the best value for the observed primordial $^4$He abundance is 0.249, then, for $\eta_{10} \sim 6$, the central value for $N_\nu$ is very close to 3. A maximum likelihood analysis on $\eta$ and $N_\nu$ based on the above $^4$He and D abundances finds the (correlated) 95% CL ranges to be $4.9 < \eta_{10} < 7.1$ and $1.8 < N_\nu < 4.5$ [90]. Identical results are obtained using a simpler method to extract such bounds based on $\chi^2$ statistics, given a set of input abundances [91].

The CMB power spectrum in the damping tail is independently sensitive to $N_\nu$ (*e.g.* [92]) . The CMB value $N_\nu^{\rm CMB}$ probes the cosmic radiation content at (re)combination, so a discrepancy would imply new physics or astrophysics. Indeed, observations by the South Pole Telescope implied $N_\nu^{\rm CMB} = 3.85 \pm 0.62$ [93], prompting discussion of "dark radiation" such as sterile neutrinos [94]. However, recent *Planck* results give $N_\nu^{\rm CMB} = 3.36^{+0.34}_{-0.32}$ with a precision now apparently better than BBN, which is quite consistent with the Standard Model [52]. If we *assume* that $\eta$ did not change between BBN and (re)combination, the constraint can be improved by including the recent D/H measurements, which yields $N_\nu = 3.28 \pm 0.28$ [36].

Just as one can use the measured helium abundance to place limits on $g_*$ [89], any changes in the strong, weak, electromagnetic, or gravitational coupling constants, arising *e.g.*, from the dynamics of new dimensions, can be similarly constrained [95], as can be any speed-up of the expansion rate in, *e.g.*, scalar-tensor theories of gravity [96].

The limits on $N_\nu$ can be translated into limits on other types of particles or particle masses that would affect the expansion rate of the Universe during nucleosynthesis. For example, consider 'sterile' neutrinos with only right-handed interactions of strength $G_{\rm R} < G_{\rm F}$. Such particles would decouple at higher temperature than (left-handed) neutrinos, so their number density ($\propto T^3$) relative to neutrinos would be reduced by any subsequent entropy release, *e.g.*, due to annihilations of massive particles that become non-relativistic between the two decoupling temperatures. Thus (relativistic) particles with less than full strength weak interactions contribute less to the energy density than particles that remain in equilibrium up to the time of nucleosynthesis [97]. If we impose $N_\nu < 4$ as an illustrative constraint, then the three right-handed neutrinos must have a temperature $3(T_{\nu_{\rm R}}/T_{\nu_{\rm L}})^4 < 1$. Since the temperature of the decoupled $\nu_{\rm R}$ is determined by entropy conservation (see Big Bang Cosmology review), $T_{\nu_{\rm R}}/T_{\nu_{\rm L}} = [(43/4)/g_*(T_{\rm d})]^{1/3} < 0.76$, where $T_{\rm d}$ is the decoupling temperature of the $\nu_{\rm R}$. This requires $g_*(T_{\rm d}) > 24$, so decoupling must have occurred at $T_{\rm d} > 140$ MeV. The decoupling temperature is related to $G_{\rm R}$ through $(G_{\rm R}/G_{\rm F})^2 \sim (T_{\rm d}/3\,{\rm MeV})^{-3}$, where 3 MeV is the decoupling temperature for $\nu_{\rm L}$s. This yields a limit $G_{\rm R} \lesssim 10^{-2} G_{\rm F}$. The above argument sets lower limits on the masses of new $Z'$ gauge bosons to which right-handed neutrinos would be coupled in models of superstrings [98], or extended technicolor [99]. Similarly a Dirac magnetic moment for neutrinos, which would allow the right-handed states to be produced through scattering and thus increase $g_*$, can be significantly constrained [100], as can any new interactions for neutrinos that have a similar effect [101]. Right-handed states can be populated directly by helicity-flip scattering if the neutrino mass is large enough, and this property has been used to infer a bound of $m_{\nu_\tau} \lesssim 1$ MeV taking $N_\nu < 4$ [102]. If there is mixing between active and sterile neutrinos then the effect on BBN is more complicated [103].

BBN limits on the cosmic expansion rate constrain supersymmetric scenarios in which the neutralino or gravitino are very light, so that they contribute to $g_*$ [104]. A gravitino in the mass range $\sim 10^{-4} - 10$ eV will affect the expansion rate of the Universe similarly to a light neutralino (which is however now probably ruled out by collider data, especially the decays of the Higgs-like boson). The net contribution to $N_\nu$ then ranges between 0.74 and 1.69, depending on the gravitino and slepton masses [105].

The limit on the expansion rate during BBN can also be translated into bounds on the mass/lifetime of non-relativistic particles that decay during BBN. This results in an even faster speed-up rate, and typically also changes the entropy [106]. If the decays include Standard Model particles, the resulting electromagnetic [107–108] and/or hadronic [109] cascades can strongly perturb the light elements, which leads to even stronger constraints. Such arguments had been applied to rule out an MeV mass for $\nu_\tau$, which decays during nucleosynthesis [110].

Decaying-particle arguments have proved very effective in probing supersymmetry. Light-element abundances generally are complementary to accelerator data in constraining SUSY parameter space, with BBN reaching to values kinematically inaccessible to the LHC. Much recent interest has focused on the case in which the next-to-lightest supersymmetric particle is metastable and decays during or after BBN. The constraints on unstable particles discussed above imply stringent bounds on the allowed abundance of such particles [109]; if the metastable particle is charged (*e.g.*, the stau), then it is possible for it to form atom-like electromagnetic bound states with nuclei, and the resulting impact on light elements can be quite complex [111]. Moreover, SUSY decays can destroy $^7$Li and/or produce $^6$Li, leading to a possible supersymmetric solution to the lithium problems noted above [112]( see [6] for a review).

These arguments impose powerful constraints on supersymmetric inflationary cosmology [108–109], particularly thermal leptogenesis [113]. These can be evaded only if the gravitino is massive enough to decay before BBN, *i.e.*, $m_{3/2} \gtrsim 50$ TeV [114]( which would be unnatural), or if it is in fact the lightest supersymmetric particle and thus stable [108,115]. Similar constraints apply to moduli – very weakly coupled fields in string theory that obtain an electroweak-scale mass from supersymmetry breaking [116].

Finally, we mention that BBN places powerful constraints on the possibility that there are new large dimensions in nature, perhaps enabling the scale of quantum gravity to be as low as the electroweak scale [117]. Thus, Standard Model fields may be localized on a 'brane,' while gravity alone propagates in the 'bulk.' It has been further noted that the new dimensions may be non-compact, even infinite [118], and the cosmology of such models has attracted considerable attention. The expansion rate in the early Universe can be significantly modified, so BBN is able to set interesting constraints on such possibilities [119].